\documentclass[final,11pt,superscriptaddress,aps,rmp,a4]{revtex4} 
\usepackage{latexsym}
\usepackage{times} 
\usepackage{graphicx}
\usepackage[tight,FIGTOPCAP]{subfigure}
\usepackage{color}
\usepackage{afterpage}

\begin{document}

\title{Universal Features in the Genome-level Evolution of Protein
  Domains}


\author{M.~Cosentino Lagomarsino} 
\affiliation{Universit\`a degli Studi di Milano, Dip.
    Fisica.  Via Celoria 16, 20133 Milano, Italy} 
  \email[ e-mail address: ]{Marco.Cosentino-Lagomarsino@unimi.it}
\altaffiliation{and I.N.F.N. Milano, Italy. Tel. +39 02 50317477 ; fax
  +39 02 50317480 }
\author{A.L.~Sellerio}
\affiliation{Universit\`a degli Studi di Milano, Dip.
    Fisica.  Via Celoria 16, 20133 Milano, Italy} 
\author{P.D.~Heijning}
\affiliation{Universit\`a degli Studi di Milano, Dip.
    Fisica.  Via Celoria 16, 20133 Milano, Italy} 
\author{B.~Bassetti}
\affiliation{Universit\`a degli Studi di Milano, Dip.
    Fisica.  Via Celoria 16, 20133 Milano, Italy} 
\altaffiliation{and I.N.F.N. Milano, Italy. Tel. +39 02 50317477 ; fax
  +39 02 50317480 }

\date{\today}

\baselineskip24pt

\maketitle

{\bf

  Protein domains are found on genomes with notable statistical
  distributions, which bear a high degree of similarity.  Previous
  work has shown how these distributions can be accounted for by
  simple models, where the main ingredients are probabilities of
  duplication, innovation, and loss of domains. However, no one so far
  has addressed the issue that these distributions follow definite
  trends depending on protein-coding genome size only.  We present a
  stochastic duplication/innovation model, falling in the class of
  so-called \emph{Chinese Restaurant Processes}, able to explain this
  feature of the data.  Using only two \emph{universal} parameters,
  related to a minimal number of domains and to the relative weight of
  innovation to duplication, the model reproduces two important
  aspects: (a) the populations of domain classes (the sets, related to
  homology classes, containing realizations of the same domain in
  different proteins) follow common power-laws whose cutoff is
  dictated by genome size, and (b) the number of domain families is
  universal and markedly sublinear in genome size. An important
  ingredient of the model is that the innovation probability decreases
  with genome size. We propose the possibility to interpret this as a
  global constraint given by the cost of expanding an increasingly
  complex interactome.  Finally, we introduce a variant of the model
  where the choice of a new domain relates to its occurrence in
  genomic data, and thus accounts for fold specificity.  Both models
  have general quantitative agreement with data from hundreds of
  genomes, which indicates the coexistence of the well-known
  specificity of proteomes with robust self-organizing phenomena
  related to the basic evolutionary ``moves'' of duplication and
  innovation.  }


%

\section*{Introduction}

The availability of many genome sequences gives us abundant
information, which is, however, very difficult to decode.  As a
consequence, in order to advance our understanding of biological
processes at the whole-cell scale, it becomes very important to
develop higher-level descriptions of the contents of a genome. At the
level of the proteome, an effective scale of description is provided
by protein domains~\cite{OT05}. Domains are the unit-shapes (or
``folds'') forming proteins~\cite{BT99}.  They constitute independent
thermodynamically stable structures.  A domain determines a set of
potential functions and interactions for the protein that carries it,
for example DNA- or protein-binding capability or catalytic
sites~\cite{KWK02,OT05}.  Therefore, domains underlie many of the
known genetic interaction networks. For example, a transcription
factor or an interacting pair of proteins need the proper binding
domains~\cite{MBT03,NBG+05}, whose binding sites define transcription
networks and protein-protein interaction networks respectively.

Protein domains are related to sets of sequences of the protein-coding
part of genomes. Multiple sequences give rise to the same shape, and
the choice of a specific sequence in this set fine-tunes the function,
activity and specificity of the inherent physico-chemical properties
that characterize a shape. A domain then defines naturally a ``domain
class'', constituted by all its realizations in the genome, or all the
proteins using that given shape for some function.
Overall, domains can be seen as an ``alphabet'' of basic elements of
the protein universe.  Understanding the usage of domains across
organisms is as important and challenging as decoding an unknown
language.
Much as the letters of linguistic alphabets, the domains observable
today are few, probably of the order of $10^5$~\cite{KWK02}.  This
number is surprisingly lower than the number of possible protein
sequences (which are in general a hundred orders of magnitude more
numerous).
In the course of evolution, domains are subject to the dynamics of
genome growth (by duplication, mutation, horizontal transfer, gene
genesis, etc.) and reshuffling (by recombination etc.), under the
constraints of selective pressure~\cite{KWK02,QLG01}.  These drives
for combinatorial rearrangement, together with the defining modular
property of domains, lead to the construction of increasingly richer
sets of proteins~\cite{RBT+04}. In other words, domains are
particularly flexible evolutionary building blocks.

In particular, the sequences of two duplicate domains that diverged
recently will be very similar, so that one can also give a strictly
evolutionary definition of protein domains~\cite{KWK02}, as regions of
a protein sequence that are highly conserved.
The (interdependent) structural and evolutionary definitions of
protein domains given above have been used to produce systematic
hierarchical taxonomies of domains that combine information about
shapes, functions and sequences~\cite{MBH+95,OMJ+97}. Generally, one
considers three layers, each of which is a subclassification of the
previous one. The top layer of the hierarchy is occupied by ``folds'',
defined by purely structural means. It is then possible, though it
seems quite rare,
that a fold is polyphyletic, i.e. found from different paths in
evolution.  The intermediate ``superfamily'' class is also mainly
defined by spatial shape, with the aid of sequence and functional
annotations to guarantee monophyly. Finally, the ``family'' class is
defined by sequence similarity.

The large-scale data stemming from this classification effort enable
to tackle the challenge of understanding the alphabet of protein
domains~\cite{RST+06,OT05,BBK+05,WBB06}. In particular, they have been
used to evaluate the laws governing the distributions of domains and
domain families~\cite{HvN98,QLG01,KWR+02,Kuz02,AD05}.  As noted by
previous investigators, these laws are notable and have a high degree
of universality.
We reviewed these observations performing our own analysis of data on
folds and superfamilies from the SUPERFAMILY database~\cite{WMV+07}
(Supplementary Note \ref{sec:fold}).  Using the number of domains
$n$\footnote{ Note that this quantity is linear in the number of
  proteins, thus the two measures of genome size are interchangeable }
to measure the size of a genome, we have the following observations,
that confirm (and in part extend) previous ones.
\begin{itemize}
\item[(i)] The number of domain families (or distinct hits of the same
  domain) concentrates around a curve $F(n)$ that is markedly
  sublinear with size (figure \ref{fig:f_n}A), perhaps saturating.
\item[(ii)] The number $F(j,n)$ of domain classes having $j$ members
  (in a genome of size $n$) follows the power-law $\sim 1/ j^{1+
    \alpha}$, where the fitted exponent $1+\alpha$ typically lies
  between 1 and 2 (figure \ref{fig:hist}).
\item[(iii)] The fitted exponent of this power-law appears to decrease
  with genome size (figure \ref{fig:hist}A), and there is evidence for
  a cutoff that increases linearly with $n$ (figure \ref{fig:hist}C). 
\end{itemize}

Recent modeling efforts focused mainly on observation (ii), or the
fact the the domain class distributions are power laws. They explored
two main directions.  First, a ``designability''
hypothesis~\cite{LTW98}, which claims that domain occurrence is due to
accessibility of shapes in sequence space.  While the debate is open,
this alone seems to be an insufficient explanation, given for example
the monophyly of most folds in the taxonomy~\cite{KWK02,DS07}.
A second, ``genome growth'' hypothesis ascribing the emergence of
power-laws to a generic preferential-attachment principle due to gene
duplication seems to be more successful. Growth models were formulated
as nonstationary, duplication-innovation
models~\cite{QLG01,KLQ+06,DS05}, and as stationary
birth-death-innovation models~\cite{KWR+02,KWK03,KWB+04,KBK05}. They
were successful in describing to a consistent quantitative extent the
observed power laws.
However, in both cases, each genome was fitted by the model with a
specific set of kinetic coefficients, governing duplication, influx of
new domain classes, or death of domains. Another approach used the
same modeling principles in terms of a network view of homology
relationships within the collective of all protein
structures~\cite{DSS02,Dok05}

On the other hand, the common trend for the number of domain classes
at given genome size and the common behavior of the observed power
laws in different organisms having the same size (observations i-iii
above), call for a unifying behavior in these distributions, which has
not been addressed so far.  Here, we first define and relate to the
data a non-stationary duplication-innovation model in the spirit of
Gerstein and coworkers~\cite{QLG01}. Compared to this work, our main
idea is that a newly added domain class is treated as a
\emph{dependent} random variable, conditioned by the preexisting
coding genome structure in terms of domain classes and number.
We will show that this model explains observations (i-iii) with a
\emph{unique} underlying stochastic process having only two
\emph{universal} parameters of simple biological interpretation, the
most important of which is related to the relative weight of
adding a domain belonging to a new family and duplicating an existing
one.  In order to reproduce the data, the innovation probability of
the model has to decrease with proteome size, i.e. such as it is less
likely to find new domains in genomes with increasingly larger number
of domains.  This feature is absent in previous models, and we suggest
the possibility to interpret it as a consequence of the computational
cost for adding a new domain class in a genome. This cost could be
associated to a rewiring in the existing regulatory interaction
networks, needed to accomodate an new domain class, correspondong to
an extra set of functions.
Finally, we show how the specificity of domain shapes, introduced in
the model using empirical data on the usage of domain classes across
genomes, can improve quantitative agreement of the model with data,
and in particular predict the saturation of the number of domain
classes $F(n)$ at large genome sizes.

\section*{Model and Results}

\subsection*{Main Model}

\begin{figure}[!hp]
  \centering
  \includegraphics[width=\textwidth]{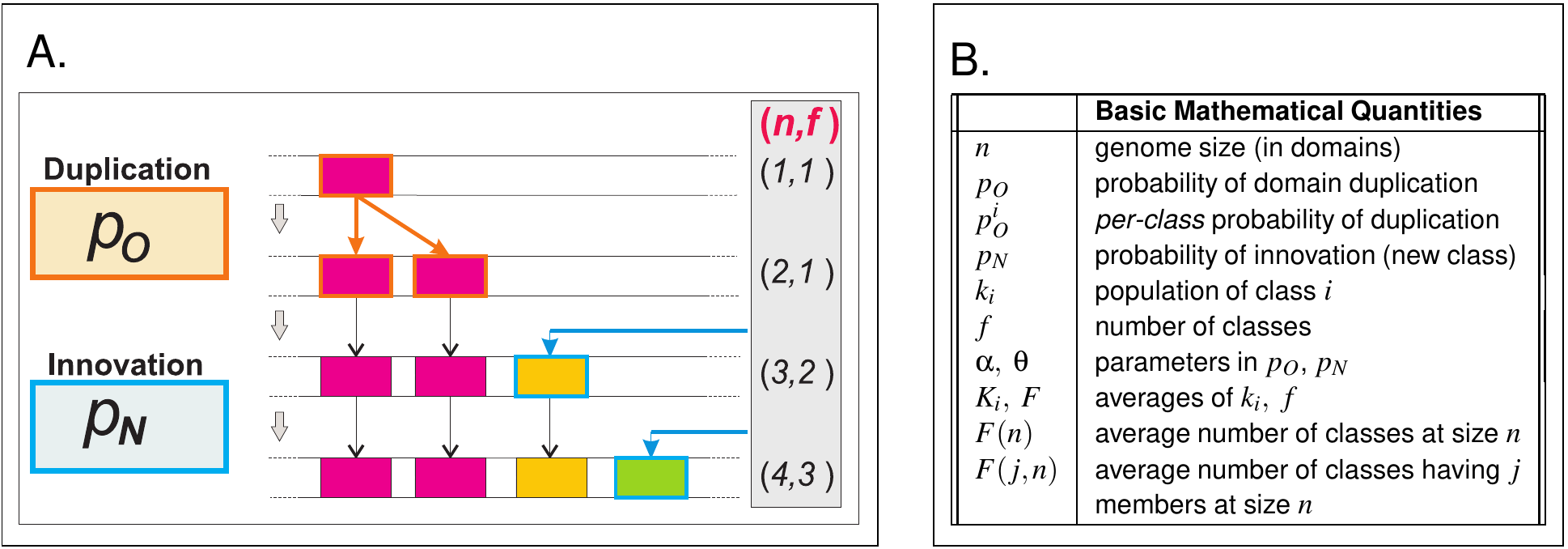}
  \caption{ Evolutionary Model.  A. Scheme of the basic moves. A
    domain of a given class (represented by its color) is duplicated
    with probability $p_N$, giving rise to a new member of the same
    family (hence filled with the same color. Alternatively, an
    innovation move creates a domain belonging to a new domain class
    (new color) with probability $p_N$.
    B. Summary of the main mathematical quantities and parameters of
    the model. 
  }
  \label{fig:intro}
\end{figure}

\paragraph*{Ingredients.}
An illustration of the model and a table resuming the main parameters
and observables are presented in figure~\ref{fig:intro}.
The basic ingredients of the model are $p_O$, the probability to
duplicate an old domain (modeling gene duplication), and $p_N$, the
probability to add a new domain class with one member (which describes
domain innovation, for example by horizontal transfer).
Iteratively, either a domain is duplicated with the former probability
or a new domain class is added with the latter.  

An important feature of the duplication move is the (null) hypothesis
that duplication of a domain has uniform probability along the genome,
and thus it is more probable to pick a domain of a larger class. This
is a common feature with previous models~\cite{QLG01}. This hypothesis
creates a ``preferential attachment'' principle, stating the fact that
duplication is more likely in a larger domain class, which, in this
model as in previous ones, is responsible for the emergence of
power-law distributions.
In mathematical terms, if the duplication probability is split as the
sum of \emph{per-class} probabilities $p_O^i$, this hypothesis
requires that $p_0^i \propto k_i$, where $k_i$ is the population of
class $i$, i.e. the probability of finding a domain of a particular
class and duplicating it is proportional to the number of members of
that class.

It is important to notice that in this model, while $n$ can be used as
an arbitrary measure of time, the weight ratio of innovation to
duplication at a given $n$ is not arbitrary, and is set by the ratio
$p_N/p_O$. In the model of Gerstein and coworkers, both probabilities,
and hence their ratio, are constant. In other words the innovation
move is considered to be statistically independent from the genome
content.  This choice has two problems.  First, it cannot give the
observed sublinear scaling of $F(n)$.  Indeed, if the probability of
adding a new domain is constant with $n$, so will be the rate of
addition, implying that this quantity will increase on average
\emph{linearly} with genome size.  It is fair to say that Gerstein and
coworkers do not consider the fact that genomes cluster around a
common curve (as shown by the data in figure \ref{fig:f_n}) and think
each of them as coming from a stochastic process with genome-specific
parameters.  Second, their choice of constant $p_N$ implies that for
larger genomes the influx of new domain families is heavily dominant
on the flux of duplicated domains. This again contradicts the data,
where additions of new domain classes are rarer with increasing genome
size.

\paragraph*{Defining Equations and Chinese Restaurant Process.}
On the contrary, motivated by the sublinear scaling of the number of
domain classes (observation (i)), we consider that $p_N$ is
conditioned by genome size. We observe (see ref. \cite{DS05}) that
constant $p_N$ makes sense thinking that new folds emerge from a
mutation process with constant rate rather than from an external
flux. This flux, coming from horizontal transfer, could be thought of
as a rare event with Poisson statistics and characteristic time
$\tau$, during which the influx of domains is $\Theta \tau$. In this
case it is immediate to verify that $f(n)$ has mean value given by  $
\sum_{j=1}^n \frac{\Theta}{\Theta+n}$ thus growing as  $\Theta \log n$.
This scenario is complementary to the one of Gerstein and coworkers
because old domain classes limits the universe that new classes can explore.

One can think of intermediate scenarios between the two. The simplest
scheme, which turns out to be quite general, implies a dependence of
$p_N$ by $n$ and $f$, where $n$ is the size (defined again as the
total number of domains) and $f$ is the number of domain classes in
the genome.  
Precisely, we consider the expressions
$$p_O^i = \frac{k_i-\alpha}{n +
  \theta}$$ \ \ , 
hence, since $p_O = \sum_i p_O^i$,
$$
p_O = \frac{n -  \alpha f}{n +\theta} \ \ ,
$$ and 
$$
p_N = \frac{\theta + \alpha f}{n +\theta} \ \ ,
$$ 
where $\theta \geq 0$ and $\alpha \in [0,1]$.  Here $\theta$ is the
parameter representing a characteristic number of domain classes
needed for the preferential attachment principle to set in, and
defines the behavior of $f(n)$ for small $n$ ( $n \rightarrow 0$).
$\alpha$ is the most important parameter, which will set the scaling
of the duplication/innovation ratio (table \ref{tab:one}),
Intuitively, for smaller $\alpha$ the process slows down the growth of
$ f $ at smaller values of $n$ ( necessarily $f < n $ ); and since
$p_N$ is asymptotically proportional to the class density $f/n$, it is
harder to add a new domain class in a larger, or more heavily
populated genome.  As we will see, this implies $p_N/p_O \rightarrow
0$ as $n \rightarrow \infty$, corresponding to an increasingly
subdominant influx of new fold classes at larger sizes. We will show
that this choice reproduces the sublinear behavior for the number of
classes and the power-law distributions described in properties
(i-iii).
%

This kind of model has previously been explored in a different context
in the mathematical literature under the name of Pitman-Yor, or
Chinese Restaurant Process (CRP)~\cite{Pit02,PY97,Ald85,Kin75}.
In the Chinese restaurant metaphor, domain realizations correspond to
customers and tables are domain classes.  A domain which is member of
a given class is a customer sitting at the corresponding table.  In a
duplication event, a new customer is seated at a table with a
preferential attachment (or packing) principle, and in an innovation
event, a new table is added.

\paragraph*{Theory and Simulation.}
We investigated this process using analytical asymptotic equations and
simulations.  The natural random variables involved in the process are
$f$, the number of tables or domain classes, $k_i$ the population of
class $i$, and $n_i$, the size at birth of class $i$.
Rigorous results for the probability distribution of the fold usage
vector $\{k_1,...,k_f\}$ confirm the results of our scaling argument.
It is important to note that in this stochastic process, large $n$
limit values of quantities such as $k_i$ and $f$ do not converge to
numbers, but rather to random variables~\cite{Pit02}.

Despite of this property, it is possible to understand the scaling of
the averages $K_i$ and $F$ (of $k_i$ and $f$ respectively) at large
$n$, writing simple ``mean-field'' equations, for continuous $n$. From
the definition of the model, we obtain $\partial_n K_i(n) =
\frac{K_i-\alpha}{n+\theta}$, and $\partial_n F(n)= \frac{\alpha
  F(n)+\theta}{n+\theta}$. These equations have to be solved with
initial conditions $K_{i}(n_i) = 1$,
and $F(0)=1$. Hence, for $\alpha \ne 0$, one
has $ K_i(n) = (1-\alpha) \frac{n+\theta}{n_i+\theta} +\theta$, and
\begin{displaymath}
  F(n)=  \frac{1}{\alpha}
  \left[(\alpha+\theta)\left(\frac{n+\theta}{\theta}\right)^{\alpha}
    -\theta \right] \sim n^{\alpha} \ \ ,
\end{displaymath}
while, for $\alpha = 0$
\begin{displaymath}
  F(n)= \theta \log (n+ \theta) \sim \log(n) \ \ .
\end{displaymath}
These results imply that the expected asymptotic scaling of $F(n)$ is
sublinear, in agreement with observation (i).
 
The mean-field solution can be used to compute the asymptotics of
$P(j,n) = F(j,n)/F(n)$~\cite{BA99}. This works as follows. From the
solution, $j> K_i(n)$ implies $n_i > n^*$, with $n^*=\frac{(1-\alpha)n
  -\theta (j-1)}{j-\alpha}$, so that the cumulative distribution can
be estimated by the ratio of the (average) number of domain classes
born before size $n^*$ and the number of classes born before size $n$,
$P(K_i(n) > j) = F(n^*)/F(n)$. $P(j,n)$ can be obtained by derivation
of this function. For $n,j \rightarrow \infty$, and $j/n$ small, we
find
\begin{displaymath}
  P(j,n)    \sim j^{-(1+\alpha)}
\end{displaymath}
for $\alpha \ne  0$, and
\begin{displaymath}
 P(j,n) \sim  \frac{\theta}{j}
\end{displaymath}
for $\alpha = 0$. The above formulas indicate that the average
asymptotic behavior of the distribution of domain class populations is
a power law with exponent between $1$ and $2$, in agreeement with
observation (ii).

The trend of the model of Gerstein and coworkers can be found for
constant $p_N,\ p_O$ and gives a linearly increasing $F(n)$ and a
power-law distribution with exponent larger than $2$ for the domain
classes.  A comparative scheme of the asymptotic results is presented
in table \ref{tab:one}.  We also verified that these results are
stable for introduction of domain loss and global duplications in the
model (see Supplementary Note \ref{sec:variants}).
Incidentally, we note that also the ``classic'' Barabasi-Albert
preferential attachment scheme~\cite{BA99} can be reproduced by a
modified model where at each step a new domain family (or new network
node) with on average $m$ members (edges of the node) is introduced,
and at the same time $m$ domains are duplicated (the edges connecting
old nodes to the new node).

Going beyond scaling, the probability distributions generated by a CRP
contain large finite-size effects that are relevant for the
experimental genome sizes.  In order to evaluate the behavior and
estimate parameter values keeping into account stochasticity and the
small system sizes, we performed direct numerical simulations of
different realizations of the stochastic process (figures
\ref{fig:f_n}B and \ref{fig:hist}B and C).
The simulations allow to measure $f(n)$, and $F(j,n)$ for finite
sizes, and in particular for values of $n$ that are comparable to
those of observed genomes. At the scales that are relevant for
empirical data, finite size corrections are substantial.  Indeed, the
asymptotic behavior is typically reached for sizes of the order of $n
\sim 10^6$, where the predictions of the mean-field theory are
confirmed.

Comparing the histogram of domain occurrence of model and data, it
becomes evident that the intrinsic cutoff set by $n$ at the causes the
observed drift in the fitted exponent described in observation (iii),
and shown in figure \ref{fig:hist}A and B.  In other words, the
observed common behavior of the slopes followed by the distribution of
domain class population for genomes of similar sizes can be ascribed
to finite-size effects of a common underlying stochastic process.  We
measured the cutoff of the distributions as the population of the
largest domain class, and verified that both model and data follow a
linear scaling (figure~\ref{fig:hist}C).  This can be expected from
the above asymptotic equations, since $K_i(n) \sim n$.

The above results show that the CRP model can reproduce the observed
qualitative trends for the domain classes and their distributions for
all genomes, with one common set of parameters, for which all random
realizations of the model lead to a similar behavior.
One further question is how quantitatively close the comparison can
be.
To answer this question, we compared data for the bacterial data sets
and models with different parameters.  Although the agreement is
reasonably good, this comparison makes it difficult to decide between
a model with $\alpha = 0$ and a model with finite (and definite)
$\alpha$: while the slope of $F(n)$ is more compatible with a model
having $\alpha=0$, the slopes of the internal power-law distribution
of domain families $P(j,n)$ and their cutoff as a function of $n$ is
in closer agreement to a CRP with $\alpha$ between 0.5 and 0.7 (figure
\ref{fig:f_n}B and Supplementary Note \ref{sec:cumul} and
\ref{sec:fold}).

\subsection*{Domain Family Identity and Model with Domain Specificity}
We have seen that the good agreement between model and data from
hundreds of genomes is universal and realization-independent .  On the
other hand, although one can clearly obtain from the basic model all
the qualitative phenomenology, the quantitative agreement is not
completely satisfactory, as both qualitative behaviors observed in the
model for $\alpha = 0$ and $\alpha > 0$ seem to agree better with only
one between the two main observables: domain distributions and
observed domain family number (figure \ref{fig:f_n} and
\ref{fig:hist}).

We will now show how a simple variant of the model that includes a
constraint based on empirically measured usage of individual domain
classes can bypass the problem, without upsetting the underlying ideas
presented above. Indeed, there exist also \emph{specific} effects, due
to the precise functional significance and interdependence of domain
classes. These give rise to correlations and trends that are clearly
visible in the data, which we have analyzed more in detail in a
parallel study~\cite{HSB+}.  Here, we will consider simply the
empirical probabilities of usage of domain families for 327 bacterial
genomes in the SUPERFAMILY database~\cite{WMV+07} (figure
\ref{fig:f_n}C). These observables are largely uneven and functional
annotations clearly show the existence of ubiquitous domain classes,
which correspond to ``core'' or vital functions, and marginal ones,
that are used for more specialized or contextual scopes~\cite{HSB+}.

In order to identify model domain classes with empirical ones, it is
necessary to label them.  We assign each of the labels a positive or
negative weight, according to its empirical frequency measured in
figure \ref{fig:f_n}C. A genome can then be assigned a cost function,
according to how much its domain family compositions resembles the
average one. In other words, the genome receives a positive score for
every ubiquitous family it uses, and a negative one for every rare
domain family.  We then introduce a variant in the basic moves of the
model, which can be thought of as a genetic algorithm. This variant
proceeds as follows.  In a first substep, the Chinese restaurant model
generates a population of candidate genome domain compositions, or
virtual moves. Subsequently, a second step discards the moves with
higher cost, i.e. were specific domain classes sre used more
differently from the average case.  Note that the virtual moves could
in principle be selected using specific criteria that keep into
account other observed features of the data than the domain family
frequency.
The model is described more in detail in Supplementary Note
\ref{sec:CRPfit}.  We mainly considered the case whith two virtual
moves, which is accessible analytically.

In the modified model, not all classes are equal. The cost function
introduces a significance to the index of the domain class, or a
colored ``tablecloth'' to the table of the Chinese restaurant. In
other words, while the probability distributions in the model are
symmetric by switch of labels in domain classes~\cite{Ald85}, this
clearly cannot be the case for the empirical case, where specific
folds fulfill specific biological functions.  We use the empirical
domain class usage to break the symmetry, and make the model more
realistic. Moreover, the labels for domain classes identify them with
empirical ones, so that the model can be effectively used as a null
model.

Simulations and analytical calculations show that this modified model
agrees very well with observed data.  Figures \ref{fig:f_n}B and
\ref{fig:hist}B show the comparison of simulations with empirical
data. The agreement is quantitative. In particular, the values of
$\alpha$ that better agree with the empirical behavior of the number
of domain classes as a function of domain size $F(n)$ are also those
that generate the best slopes in the internal usage histograms
$F(j,n)$.  Namely, the best $\alpha$ are between 0.5 and 0.7.
Furthermore, the cost  function generates a critical value of $n$, above
which $F(n)$, the total number of domain families, becomes flat. This
behavior agrees with the empirical data better than the asymptotically
growing laws of the standard CRP model. A mean-field calculation of
the same style as the one presented above predicts the existence of
this plateau (see Supplementary Note \ref{sec:CRPfit}).

\section*{Discussion and Conclusions}

The model shows that the observed common features in the number and
population of domain classes organisms with similar proteome sizes can
be explained by the basic evolutionary moves of innovation and
duplication.  This behavior can be divided into two distinct universal
features.  The first is the common scaling with genome size of the
power laws representing the population distribution of domain classes
in a genome. This was reported early on by Huynen and van
Nimwegen~\cite{HvN98}, but was not considered by previous models.  The
second feature is the number of domain families versus genome size
$F(n)$, which clearly shows that genomes tend to cluster on a common
curve. This fact is remarkable, and extends previous observations. For
example, while it is known that generally in bacteria horizontal
transfer is more widespread than in eukaryotes, the common behavior of
innovation and duplication depending on coding genome size only might
be unexpected.
The sublinear growth of number of domain families with genome size
implies that addition of new domains is conditioned to genome size,
and in particular that additions are rarer with increasing size.
%

Previous literature on modeling of large-scale domain usage
concentrated on reproducing the observed power-law behavior and did
not consider the above-described common trends.  In order to explain
these trends we introduce a size dependency in the ratio of innovation
to duplication $p_N/p_O$. This feature is absent in the model of
Gerstein and coworkers, which is the closest to our formalism. We have
shown that this choice is generally due to the fact that $p_N$ is
conditioned by genome size.  Furthermore, we can argue on technical
grounds that the choice of having constant $p_O$ and $p_N$ would be
more artificial, as follows.  If one had $p_0^i = k_i / n$, the total
probability $p_O$ would be one, since the total population $n$ is the
sum of the class populations $k_i$, and there would not be
innovation. In order to build up an innovation move, and thus $p_N >
0$, one has to subtract small ``bits'' of probability from $p_O^i$.
If $p_N$ has to be constant, the necessary choice is to take $p_O^i =
k_i / n - p_N/f$, where $f$ is the number of domain classes in the
genome.  This means that the probability of duplication for a member
of one class would be awkwardly dependent on the total number of
classes.
Furthermore, we have addressed the role of specificity of domain
classes, by considering a second model where each class has a specific
identity, given by its empirical occurence in the genomes of the
SUPERFAMILY data set. This model, which gives up the complete symmetry
of domain classes, gives the best quantitative agreement with the
data, and is a good candidate for a null model designed for
genome-scale studies of protein domains.  More specific biological and
physical properties, such as individual domain function and
designability~\cite{DS07,ZCS+07,OT05} come in at the more detailed
level of description of how domains are actually used to form
functional proteins.

It is useful to spend a few words on the role of common ancestry in
these observations. Clearly, empirical genomes come from
intertwined evolutionary paths, which determines their current
states.  On the other hand, the probability distributions predicted by
the model are essentially the same for all histories (at fixed
parameters), which can be taken as an idication that for these
observations, the basic moves are more important than the shared
evolutionary history.
While the scaling laws are found independently on the realization of
the Chinese restaurant model, the uneven presence of domain classes
can be seen as strongly dependent on common evolutionary history.
Averaging over independent realizations, the prediction of the
standard model would be that the frequency of occurrence of any domain
class would be equal, as no class is assigned a specific label. In the
Chinese restaurant metaphor, the customers only choose the tables on
the basis of their population, and all the tables are equal for any
other feature. However, if one considers a single realization, which
is an extreme, but comparatively more realistic description of common
ancestry, the classes that appear first are obviously more common
among the genomes. In the specific variant, the empirically ubiquitous
classes are given a lower cost function, and tend to appear first in
all realizations.

\begin{figure}[!hp]
  \centering
  \includegraphics[width=\textwidth]{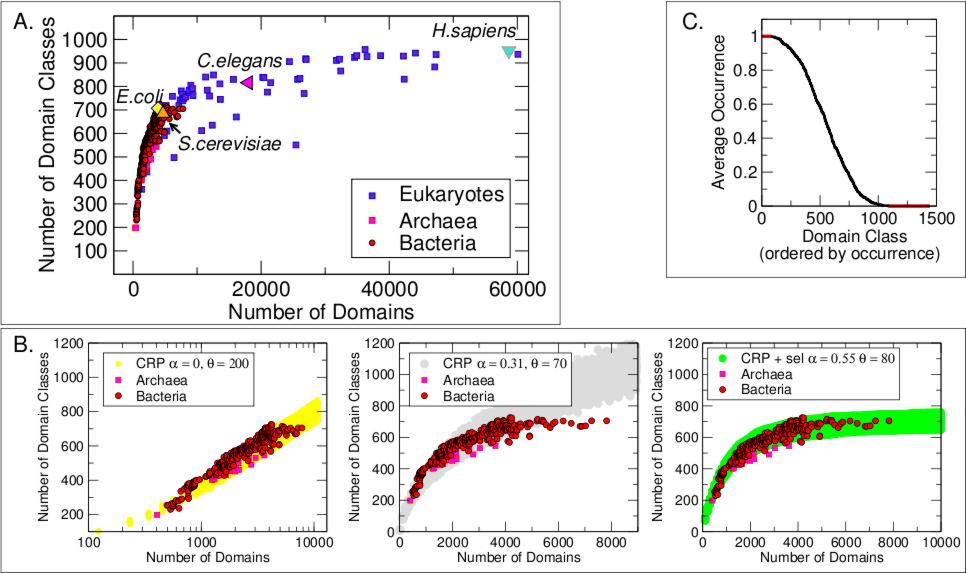}
  \caption{ 
    Number of domain classes versus genome size.  A. Plot of
    empirical data for 327 prokaryotes, 75 eukaryotes, and 27 archaeal
    genomes. Data refer to superfamily domain classes from the
    SUPERFAMILY database~\cite{WMV+07}. Larger data points indicate
    specific examples.  Data on SCOP folds follow the same trend
    (Supplementary Note \ref{sec:fold}).  B. Comparison of data on
    prokaryotes (red circles) with simulations of 500 realizations of
    different variants of the model (yellow, grey, and green shade in
    the different panels), for fixed parameter values. Data on archaea
    are shown as squares.  $\alpha=0$ (left panel, graph in log-linear scale) 
    gives a trend that is more compatible with the observed scaling
    than $\alpha>0$ (mid panel).  However, the empirical distribution
    of folds in classes is quantitatively more in agreement with
    $\alpha>0$ (see table \ref{tab:one} and figure
    \ref{fig:hist}). The model that breaks the symmetry between domain
    classes and includes specific selection of
    domain classes (right panel) predicts a saturation of this
    curve even for high values of $\alpha$, resolving this
    quantitative conflict.
    C. Usage profile of SUPERFAMILY domain classes in prokaryotes,
    used to generate the cost function in the model with specificity. In
    the x-axis, domain families are ordered by the fraction of genomes
    they occur in. The y-axis reports their occurrence fraction. The
    red lines indicate occurrence in all or none of the prokarotic
    genomes of the data set.  }
  \label{fig:f_n}
\end{figure}

\begin{figure}[p]
  \centering
  \includegraphics[width=\textwidth]{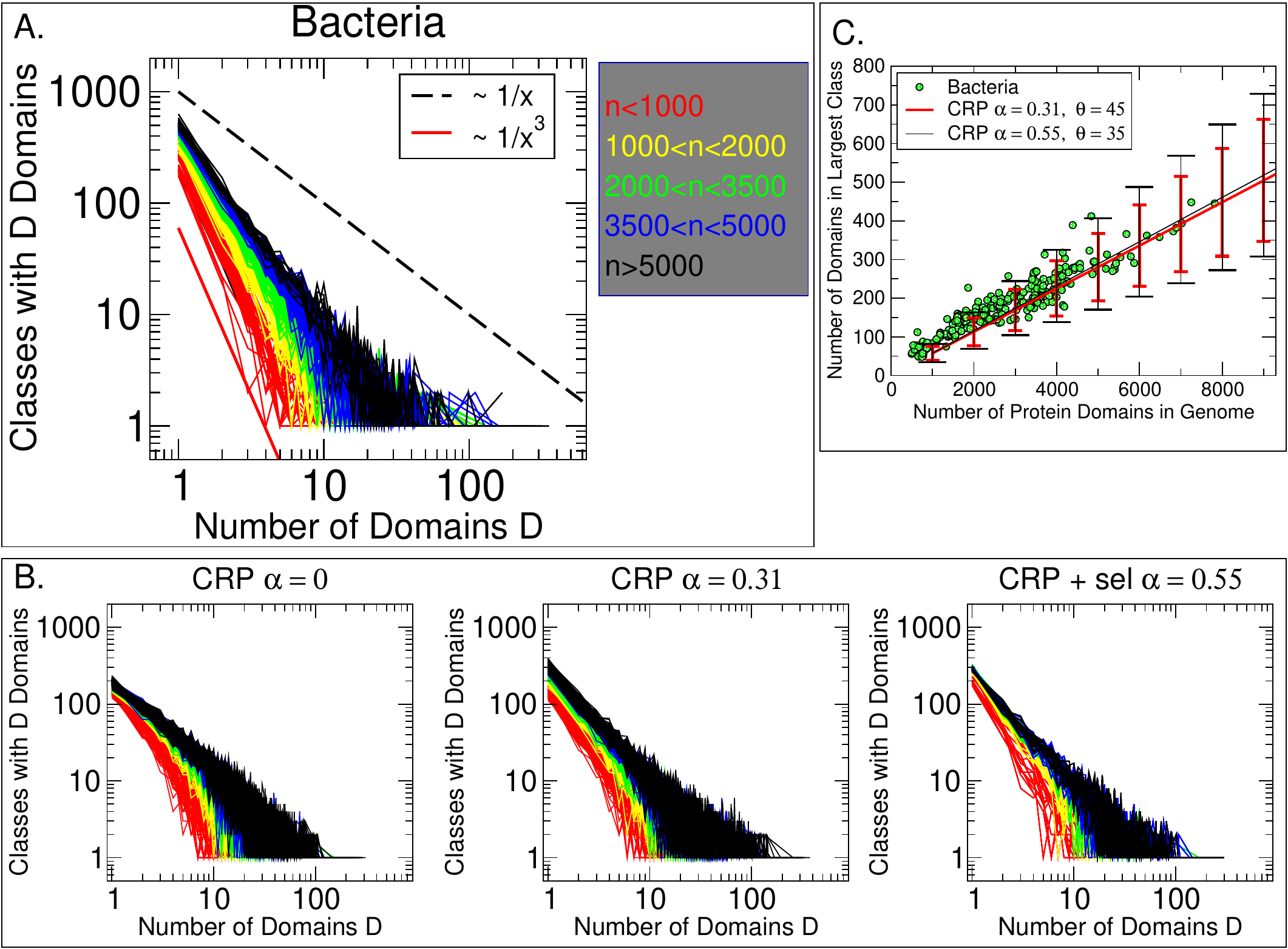}
  \caption{Internal usage of domains.
A. Histograms of domain usage; empirical data for 327 prokaryotes. The
x-axis indicates the population of a domain class, and the y-axis
reports the number of classes having a 
given population of domains.  Each of the 327 curves is a histogram
referring to a different genome. The genome sizes are color-coded as
indicated by the legend on the right. Larger genomes (black) tend to have a
slower decay, or a larger cutoff, compared to smaller genomes
(red). The continuous (red) and dashed (black) lines indicate a decay
exponent of 3 and 1 respectively.  
B. Histograms of domain usage for 50 realizations of the model at
genome sizes between 500 and 8000. The color code is the same as
in panel A. All data are in qualitative agreement with the empirical
ones. However, data at $\alpha=0$ appear to have a faster decay compared
to empirical data. This is also evident looking at the cumulative distributions
(Supplementary Note \ref{sec:cumul}). 
The right panel refers to the model with specificity,
at parameters values that reproduce well the empirical number of
domain classes at a given genome size (figure \ref{fig:f_n}). 
C. Population of the maximally populated domain class as a function of
genome size. Empirical data of prokaryotes (green circles), are
compared to realizations of the CRP, for two different values of
$\alpha$, the lines indicate  averages over 500 realizations,
with error bars indicating standard deviation. $\alpha=0$ 
can reproduce the empirical trend only qualitatively (not shown).  
Data from the SUPERFAMILY database\cite{WMV+07}.  }
  \label{fig:hist}
\end{figure}

The next question worth discussing is the possible biological
interpretation of the scaling of innovation to duplication, $p_N/p_O$
as a function of proteome size $n$.  As we have shown, this ratio must
scale in the correct way with $n$ in order to reproduce the data. 
As shown in table \ref{tab:one} and in figures~\ref{fig:f_n} and
\ref{fig:hist}, this is set by the parameter $\alpha$ of the
model. Precisely, the ratio $p_N/p_O$ decreases like $\sim
n^{\alpha-1}$.  In other wirds necessarily something affects the
addition of domains with new structures relative to domains with old
structures, making it sparser with increasing size. This fact is not a
prediction of the model, but rather a feature of the data, which
constrains the model.  Note that innovation events can have the three
basic interpretations of horizontal transfers carrying new domain
classes, gene-genesis or splitting of domain classes when internal
structures diverge greatly, while duplication events can be
interpreted as real duplication, or horizontal transfers carrying
domains that belong to domain classes already present in the
genome. While this might be confusing if one focuses on the genome, it
seems reasonable to associate these processes to true ``innovations''
and ``duplications'' at the protein level.  At least for bacteria,
innovation by horizontal transfer could be the most likely event. In
this case, the question could be reduced to asking why the relative
rate of horizontal transfer of exogenous domain classes decreases with
proteome size relatively to the sum of duplication and horizontal
transfer of endogenous domain classes.

In order for $p_N/p_O$ to decrease with $n$, either $p_O$ has to
increase, or $p_N$ has to decrease, or both.
A possible source of increase of $p_0$ with $n$ is the effective
population size. Recent studies~\cite{LC03b} suggest that coding
genome size correlates with population size, and in turn this results
in reduced selective pressure, allowing the evolution of larger
genomes. Thus one can imagine that the ease to produce new
duplications and proteome size are expected to correlate, purely on
population genetics grounds.
A naive reason for the innovation probability to decrease would be
that the pool of total available domain shapes is small, which would
affect the   innovations at increasing size, while duplications
are free of this constraint.
However, this would imply that the currently observed genomes are
already at the limit of their capabilities in terms of producing new
protein shapes, while the current knowledge of protein folding does
not seem to indicate this fact. On the other hand, the limited
availability of domain classes could be true within a certain
environment, where the total pool of domain families is restricted.
We cannot exclude that the same kind of bias could be due to technical
problems in the recognition and classification of new shapes in the
process of producing the data on structural domains.  If recognition
algorithms tend to project shapes that are distinct on known ones,
they could classify new shapes as old ones with a rate that increases
with proteome size, leading to the observed scaling.
%
Finally, we would like to suggest that a reason for $p_N$ to decrease
with $n$ could be the cost for ``wiring'' new domains into existing
interaction networks.  The argument is related to the so-called
``complexity hypothesis'' for horizontal
transfers~\cite{JRL99,Ari05,LP08,WLG07}, which roughly states that the
facility for a transferred gene to be incorporated depends on its
position and status in the regulatory networks of the cell.
We suppose that, given a genome with $n$ domains (or for simplicity
monodomain genes) and $F$ domain families, the process leading to the
acceptance of a new domain family, and thus to a new class of
functions, will need a readaptation of the population of all the
domain families causing an increase $\delta n$ in the number of genes.
This increase is due to an underlying optimization problem that has to
adapt the new functions exploited by the acquired family to the
existing ones (by rewiring and expanding different interaction
networks). To state it another way, we imagine that in order to add
$\delta F$ new domain classes, or ``functions'', it is necessary to
insert $\delta n$ new degrees of freedom (``genes'') to be able to
dispose of the functions.
Now, generically, the computational cost for this optimization problem
(which, conceptually, may be regarded as a measure of the evolvability
of the system) could be a constant function of the size (and thus
$\delta n \sim \delta F $), or else polynomial or exponential in $F$
(i.e. $\delta n \sim F^{d} \delta F $, where $d$ is some positive
exponent, or $\delta n \sim \exp(F)\delta F $ respectively).
Integrating these relations gives $n \sim F$ in the first case, $n
\sim F^{d+1}$ in the second, and $n \sim \exp(F) $ in the third.
Inverting these expressions shows that the first choice leads to the
linear scaling of the model of Gerstein and coworkers, while the
second two correspond to the CRP, and to a sublinear $F(n)$, which
could follow a power law or logarithmic, depending on the
computational cost.
In other words, following this argument, accepting a new domain family
becomes less likely with increasing number of already available domain
families, as a consequence of a global constraint. This constraint
comes from the trade-off between the advantage of incorporating new
functions and the energetic or computational cost to govern them (both
of which are related to selective pressure). This hypothesis could be
tested by evaluating the rates of horizontal trasfers carrying new
domain classes in an extensive phylogenetic analysis.

In conclusion, model and data together indicate that evolution acts
conservatively on domain families, and show increasing preference with
genome size to exploiting available shapes rather than adding new
ones.
A final point can be made regarding the number of observed domains.
The model assumes that the new domain classes are drawn from an
infinite family of shapes, which can be even continuous~\cite{Pit02},
and leads to a discrete and small number of classes at the relevant
sizes.
Although physical considerations point to the existence of a small
``menu'' of shapes available to proteins~\cite{BM07}, the validity of
our model would imply that the empirical observation of a small number
of folds in nature does not count as evidence for this thermodynamic
property of proteins, but may have been a simple consequence of
evolution.

\vspace{1cm}

We thank S.~Maslov, H.~Isambert, F.~Bassetti, S.~Teichmann, 
M.~Babu, N.~Kashtan and L.D.~Hurst for helpful discussions. 


\newpage

\begin{table}[!h]
  \centering
{\large
\begin{displaymath}
\left.
   \begin{array}{c||l|l|l||l|l}
     & K_i  &\frac{p_N}{p_O}&\frac{p_N}{p_O^i}  &
     \textcolor[rgb]{1,0,0} {F(n)}  & \textcolor[rgb]{1,0,0}{ F(j,n)/F(n)}  \\ 
     \hline
     \textrm{CRP }\alpha=0 
     &  \sim n & \sim  n^{-1} & \sim  n^{-1} & 
     \textcolor[rgb]{1,0,0}{ \sim  \log(n) } & 
     \textcolor[rgb]{1,0,0}{ \sim \frac{\theta}{j}}  \\ 
     \textrm{CRP }\alpha > 0 
     &  \sim n & \sim  n^{\alpha-1} & \sim  n^{\alpha-1} 
     &  \textcolor[rgb]{1,0,0}{\sim n^{\alpha}}
     &  \textcolor[rgb]{1,0,0}{\sim j^{-(1+\alpha)}}   \\
     \textrm{Qian \emph{et al.} }
     & \sim n^{p_O} & = R  & \sim n^{1-p_O}   &   
      \textcolor[rgb]{1,0,0}{ \sim n} & 
      \textcolor[rgb]{1,0,0}{ \sim j^{-(2+R)}} \\
    \end{array} \right.
\end{displaymath}
}
\caption{Salient features of the proposed model in terms of scaling
  of the number of domain classes, compared to the model of Gerstein
  and coworkers~\cite{QLG01,KLQ+06}. The first three columns indicate
  the resulting average population of a class $K_i$, and the ratios of the
  probability to add a new class $p_N$ to the total and \emph{per-class}
  probabilities of duplication, as a function of genome size $n$. These
  latter two quantities are   asymptotically zero in the CRP, while
  they are constant or infinite in the model of Gerstein and
  coworkers. The last two columns indicate the resulting scaling of
  number of domain classes $F(n)$ and fraction of classes with $j$
  domains $F(j,n)/F(n)$. The results of the CRP agree qualitatively
  with observations (i-iii) in the text.}
  \label{tab:one}
\end{table}

\newpage

\afterpage{\clearpage}

\newpage

\afterpage{\clearpage}

\afterpage{\clearpage}

\newpage

\section*{SUPPLEMENTARY NOTES}

\renewcommand{\thesection}{S\arabic{section}}
\setcounter{figure}{0} 
\setcounter{section}{0} 
\renewcommand{\figurename}{Supporting Figure}
\renewcommand{\thefigure}{\thesection.\arabic{figure}}

\section{Cumulative Distributions for the Internal Usage of Domains}
\label{sec:cumul}

\begin{figure}[htbp]
  \centering \includegraphics[width=0.6\textwidth]{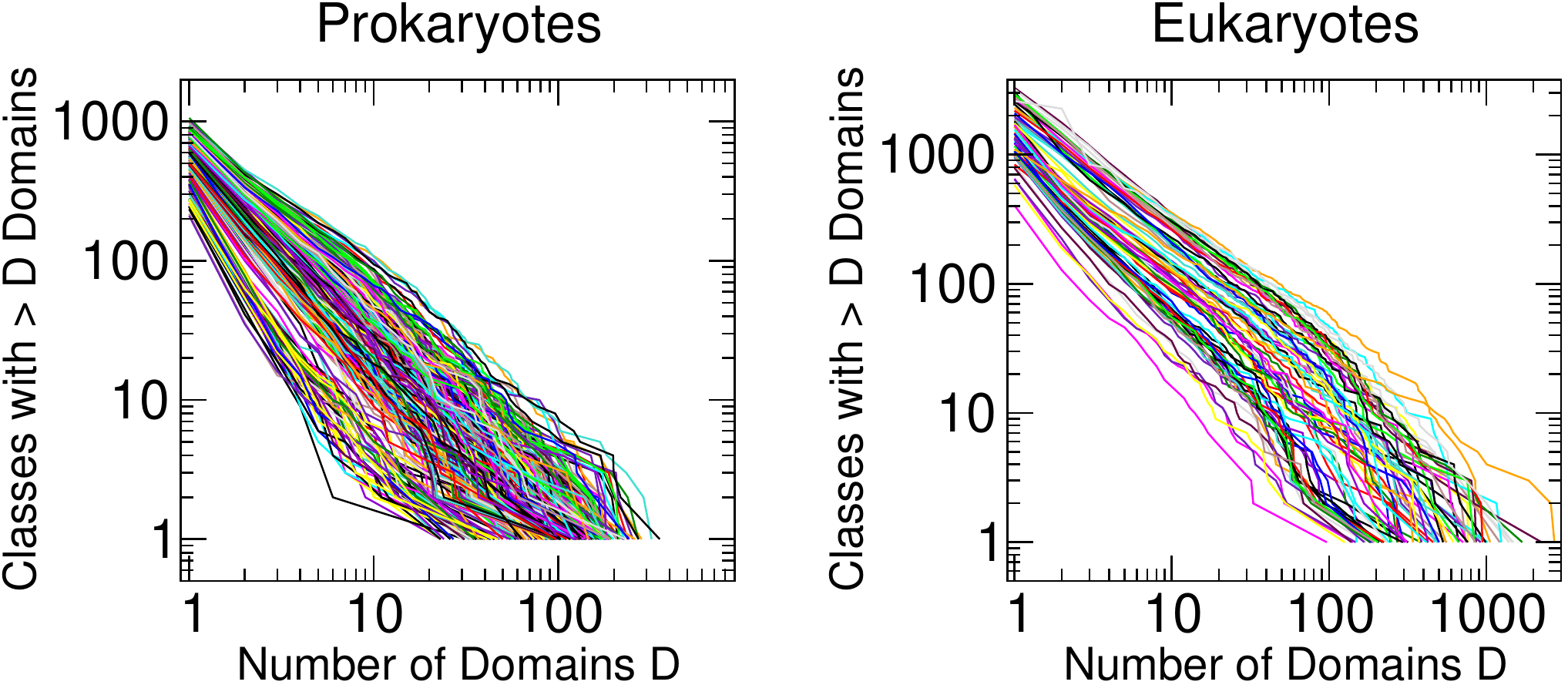}
  \caption{Empirical cumulative distributions of domain usage for
    domain classes of the SUPERFAMILY database. The x-axis reports
    domain class sizes in number of domains $D$ while the y-axis
    refers to the histogram of the number of domain classes containing
    more than $D$ domains.  The left panel is based on the same data
    on the 327 prokaryotes of figure 2A in the main text. The right
    panel refers to the 75 eukaryotes in the data set. The genome
    sizes are not color-coded to show individual plots. }
\label{fig:cum_emp}
\end{figure}

\begin{figure}[htbp]
  \centering \includegraphics[width=0.6\textwidth]{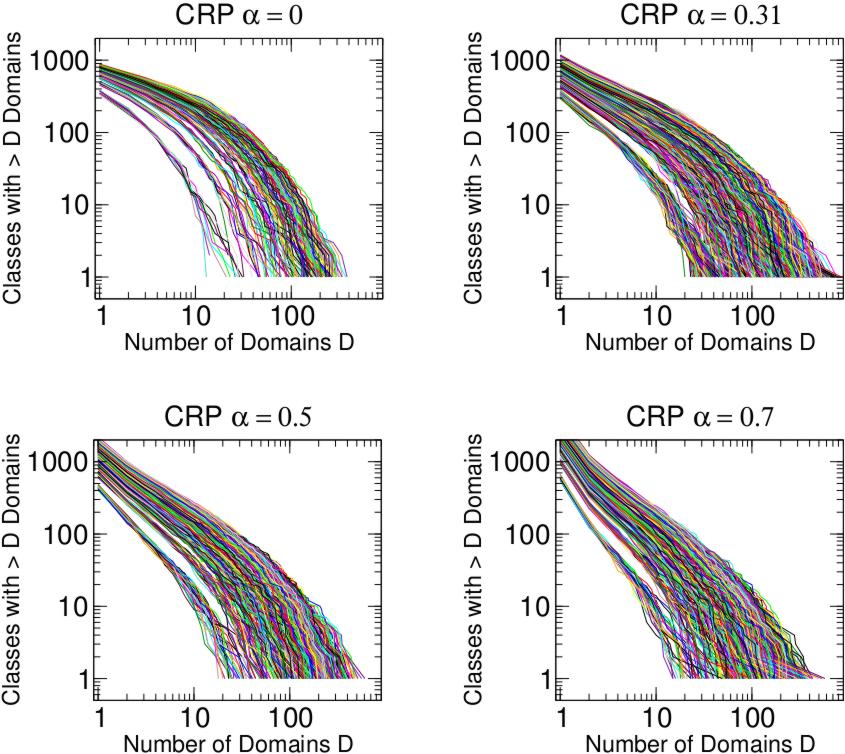}
  \caption{Cumulative histograms of domain usage for 50 realizations
    of the CRP at genome sizes between 500 and 8000. Increasing values
    of $\alpha$ are plotted in lexicographic order.}
\label{fig:cumcrp}
\end{figure}

\begin{figure}[htbp]
  \centering \includegraphics[width=0.6\textwidth]{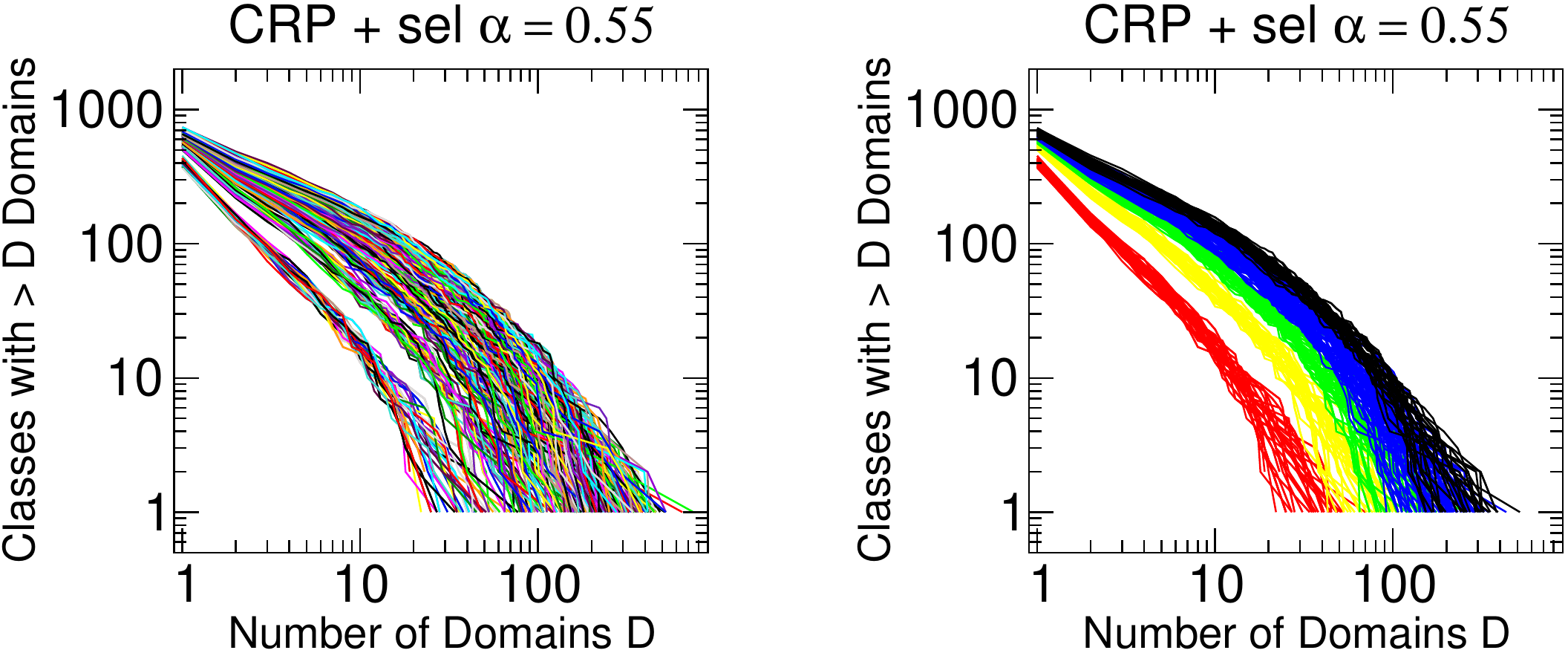}
  \caption{Cumulative histograms of domain usage for 100 realizations
    of the CRP with specific classes at genome sizes between 1000 and 8000.
    In this size range the model variant produces essentially identical
    distributions to the conventional CRP, with better agreement on
    the growth in terms of domain classes (see section
    \ref{sec:CRPfit}). The left panel is color-coded as figure 2B of
    the main text.}
\label{fig:cumcrpsel}
\end{figure}

This section briefly discusses the cumulative histograms of domain
usage for data and models. Figure \ref{fig:cum_emp} confirms the
markedly power-law behavior observed for the histograms and predicted
by the model. Comparison with the predictions of the CRP model (figure
\ref{fig:cumcrp}) shows faster decay for $\alpha = 0$. While in good
agreement with the observed number of domain classes with increasing
size (figure 1B), this parameter choice is unsatisfactory on the
quantitative side for the domain distribution in classes. This
feature, already visible in figure 2B of the main text, is even more
marked from the cumulative histograms. Better-fitting values are in
the range $\alpha = 0.5-0.7$. The CRP with specific domain classes (figure
\ref{fig:cumcrpsel}) has the same qualitative behavior as the standard
model for the distributions, while fitting well the scaling of the
classes of higher values of $\alpha$ (figure 1B and section
\ref{sec:CRPfit} below).

\section{Results for Fold Domain Classes}
\label{sec:fold}

\begin{figure}[htbp]
  \centering \includegraphics[width=0.7\textwidth]{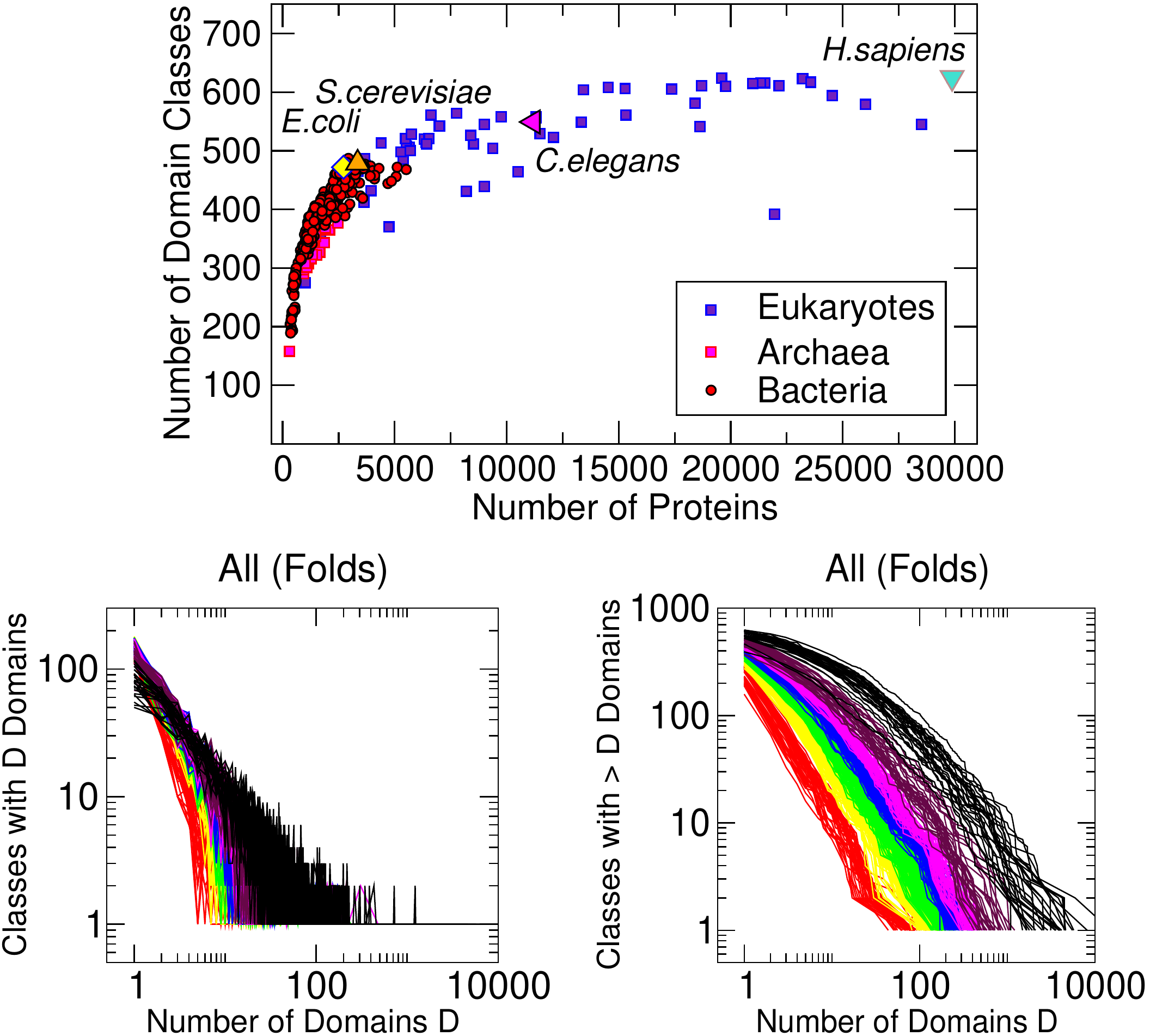} 
  \caption{Top: Number of fold classes versus genome size, the plot is
    equivalent to figure 1A, except that the x-axis reports number of
    proteins scored in the genome, rather than genome size in
    domains. Since these two quantities are quite markedly linearly
    related, the two plots are equivalent.  Bottom: histogram (left
    panel) and cumulative histogram (right panel) of domain classes
    for all genomes in the data set (eukaryotes, prokaryotes and
    archaea).}
  \label{fig:folds}
\end{figure}

\begin{figure}[htbp]
  \centering \includegraphics[width=0.4\textwidth]{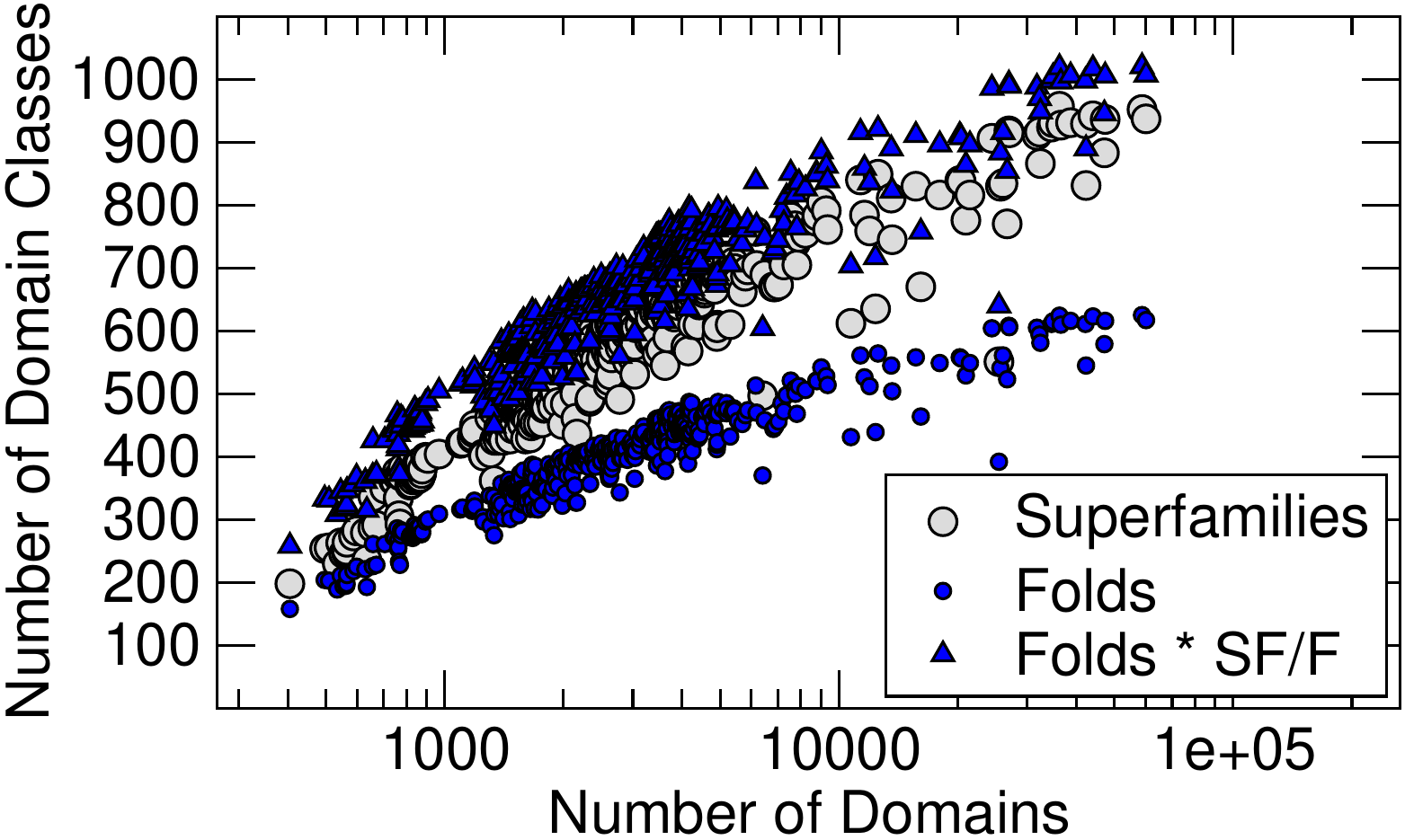} 
  \caption{Comparison of the scaling of folds and superfamilies plot
    as a function of genome size. The plots refer to all genomes in
    the SUPERFAMILY database. The plot for folds (blue small circles)
    overlaps quite well with the plot for superfamily (large grey
    circles) when multiplied by the ratio of the total number of
    domain classes in the two taxonomies (1443/884).}
  \label{fig:foldcomp}
\end{figure}

All data shown in the main text refer to the superfamily taxonomy
level, and come from the SUPERFAMILY database.  In this section, we
report the results of the same analysis in terms of SCOP folds, which
show that this category has essentially the same behavior as the
previous one (figure \ref{fig:folds}). While by definition there are
more superfamilies than folds, the number of domain classes versus
genome size has very similar scaling in the two cases.  The two plots
collapse almost exactly, when folds are rescaled by the ratio
(1443/884) of superfamilies per folds (\ref{fig:foldcomp}).
Furthermore, power-law fits of the experimental data for prokaryotes
yield an exponent $\alpha$ between 0.3 and 0.4 for both categories,
and logarithmic fits are also in agreement.

\section{CRP Model with Specific Domain Classes and Analitycal Mean Field Equations
}
\label{sec:CRPfit}

In this section we discuss the variant of the CRP model introduced in
the main text and its analytical treatment.  We first give some more
details on the definition of the model.  Generically, we consider the
following genetic algorithm. For each genome size $n$, the configuration is a
set of $M$ genomes $\{g_1(n),...,g_M(n)\}$, where each genome is a set
of $D$ domain classes populated by some domains. An iteration is
divided into two steps. A first ``proliferation'' step generates $qM$
genomes, where $q$ is a positive integer,
$\{g^{'}_1(n),...,g^{'}_{qM}(n)\}$, using the standard CRP move. A
second ``selection'' step discards the $(q-1)M$ individuals with
higher cost.


\begin{figure}[htbp]
  \centering \includegraphics[width=0.6\textwidth]{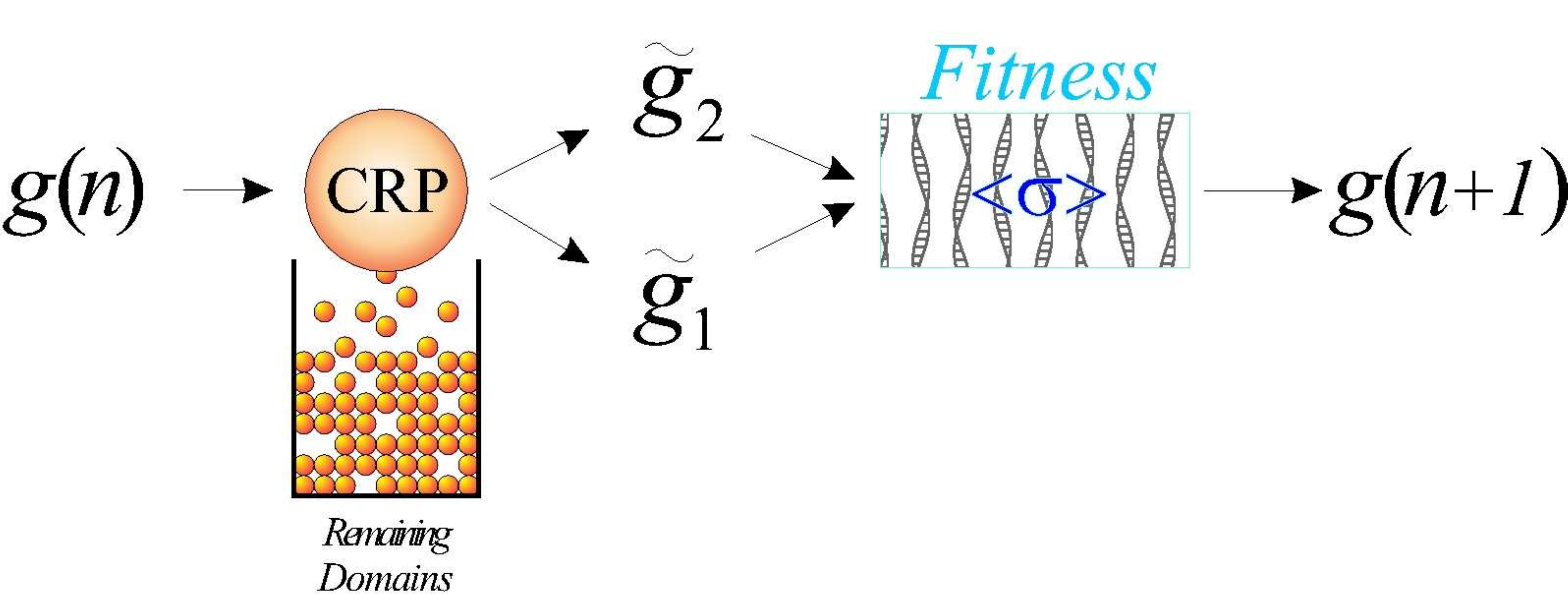}
  \caption{Scheme of the CRP variant with domain specificity. At size $n$,
    multiple (two in the figure) ``virtual'' moves are generated with
    a standard CRP model, at fixed parameters. Subsequently, the moves
    with lowest cost (one in this case) are selected. In our case,
    the cost function is chosen by comparing the domain usage of
    the model genome with the empirical usage of specific domain
    families}
\label{fig:schema_crp_fit}
\end{figure}

The cost function, for a generic
model genome $g$, can be a function $\mathcal{F}(g)$, that takes into
account some phenomenological features observed in the data. We choose
to include in $\mathcal{F}$ a minimal amount of empirical information on the
occurrence of each domain class contained in figure 1C. In other
words, we distinguish between ``universal'' domain classes, used in
most of the genomes, and ``contextual'' ones, occurring only in a few
examples. As discussed in the main text, this is sufficient to obtain
quantitative agreement with the observed domain distributions (figures
1B and 2B), which are not given to the model as an input. If domain
classes are indexed by $i=1..D$ ($D = 1443$ for Superfamilies), we define
the variable $\sigma^g_i$ as follows
\begin{displaymath}
  \sigma^l_i = \left \{  \begin{array}{cc} 
1 & \textrm{if domain class } i \textrm{ is present in genome } g \\
-1 & \textrm{if domain class } i \textrm{ is absent in genome } g \\
\end{array} \right . \  .
\end{displaymath}
The cost function of that genome is then defined as
\begin{displaymath}
  \mathcal{F}(g) = \exp \left( \sum_{i=1}^D  \sigma^g_i \langle
  \sigma^{\mathrm{EMP}}_i \rangle \right) \ \ ,
\end{displaymath}
where $\langle \sigma^{\mathrm{EMP}}_i \rangle$ is the empirical
average of the same observable:
\begin{displaymath}
  \langle \sigma^{\mathrm{EMP}}_i \rangle = \frac{1}{G} \sum_{g=1}^G
  \sigma^{g,\mathrm{EMP}}_i \ \ . 
\end{displaymath}
In the above formula $G$ is the number of observed genomes in the data
set. For example, in the case of prokaryotes in the SUPERFAMILY
database, $G=327$ and, calling $\Xi_i$ the function plotted in figure
1C, we have simply $ \langle \sigma^{\mathrm{EMP}}_i \rangle = 2 \Xi -
1$. 

\vspace{0.5cm}

For the analitycal treatment, we considered the case $M=1$, $q=2$,
where at each iteration, one genome is selected from a population of
two.  Starting from configuration $g(n)$, in the proliferation step
genomes $g^{'},g^{''}$ are generated with CRP rules, and the selection
step chooses $g(n+1) =
\mathrm{argmax}(\mathcal{F}(g^{'}),\mathcal{F}(g^{''}))$. In this
case, since the selection rule chooses strictly the maximum, it is
able to distinguish the sign of $\langle \sigma^{\mathrm{EMP}}_i
\rangle$ only. For this reason, it is sufficient to account for the
positivity (which we label by ``+'') and negativity (``-'') of this
function for a given domain index $i$. The genomes $g^{'}$ and
$g^{''}$ proposed by the CRP proliferation step can have the same
(labeled by ``$1$''), lower (``$1_+$'') or higher (``$1_-$'') cost 
than their parent, depending on $p_O$, $p_N$ and by the probabilities
to draw a universal or contextual domain family, $p_+$ and $p_-$
respectively.  Using these labels, the scheme of the possible states
and their outcome in the selection step is given by the table below.

\begin{displaymath}
  \begin{array}{|c|c|c|} 
    \hline
    \textrm{proliferation } (g^{'},g^{''})
    &\textrm{probability}&\textrm{selection}\\ 
    \hline
    (1,1)& p_O^2& \mathrm{old}\\
    (1,1_{-})& 2\: p_O\:p_N\:p_-& \mathrm{old}\\
    \hline
    (1,1_{+})& 2\: p_O\:p_N\:p_+& \mathrm{new+}\\
    (1_{+},1_{+})&p_N^2\:p_+^2)& \mathrm{new+}\\
    (1_{+},1_{-})& 2\:p_N^2\:p_-\:p_+&\mathrm{new+}\\
    \hline
    (1_{-},1_{-})&p_N^2\:p_-^2& \mathrm{new-}\\
    \hline
\end{array} 
\end{displaymath}

From this table, it is straightforward to derive the modified probabilities
$\hat p_O$ and $\hat p_N$ of the complete iteration:
\begin{displaymath}
  \hat p_O= p_O \:(p_O+ 2\:p_N p_-)
\end{displaymath}
\begin{displaymath}
  \hat p_N= p_N \:(p_N+ 2\:p_O p_+) = p_{N+} + p_{N-} \ \ ,
\end{displaymath}
where $ p_{N+} = p_N p_+ (2 - p_N p_+)$ and $ p_{N-} = p_N^2 (1 - p_+)^2 $ 
are the probabilities that the new domain is drawn from the universal
or contextual families respectively. 

We now write the macroscopic evolution equation for the number of
domain families using the same procedure as in the main text. Calling
$k^+(n)$ and $k^-(n)$ the number of domain classes that have positive
or negative $ \langle \sigma^{\mathrm{EMP}}_i \rangle$ and are
\emph{not} represented in $g(n)$,

\begin{displaymath}
 \left \{  \begin{array}{cc} 
 \partial_n F(n)=& \hat p_N\\
\partial_n k^{+}(n)=& - \hat p_{N+}\\
\partial_n k^{-}(n)=& - \hat p_{N-}\\
\end{array} \right . \ .
\end{displaymath}

Now, $p_+ = k^+ / (k^- + k^+) = k^+ / (D - F(n)$, so that we can
rewrite 

\begin{equation}
 \left \{  \begin{array}{cc}
 \partial_n F(n)=& (\frac{\alpha F(n)+\theta}{n+\theta})\:\Big[
\frac{\alpha F(n)+\theta}{n+\theta}\:+ \: \frac{2
  k^{+}(n)}{D-F(n)}(\frac{n-\alpha F(n)}{n+\theta})\Big]\\ 
\partial_n k^{+}(n)=&-(\frac{\alpha F(n)+\theta}{n+\theta})\:  \frac{
  k^{+}(t)}{D-F(n)}\:\Big[2-\: (\frac{\alpha F(n)+\theta}{n+\theta}) 
 \frac{k^{+}(n)}{D-F(n)}\Big]\\
\partial_n k^{-}(n)=& - (\frac{\alpha F(n)+\theta}{n+\theta})^2\:
(\frac{ k^{+}(n)}{D-F(n)})^2 
\end{array} \right .
\label{eq:din}
\end{equation}

\begin{figure}[htbp]
  \centering \includegraphics[width=0.7\textwidth]{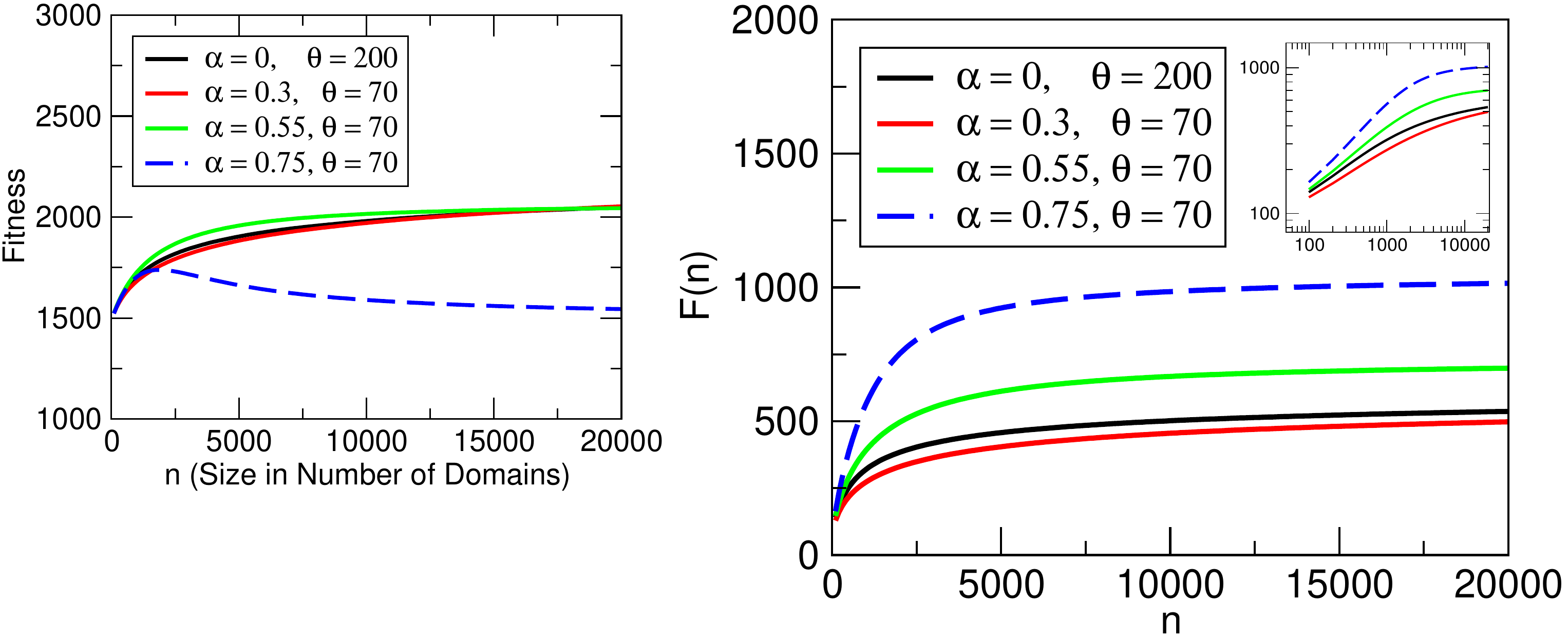} 
  \caption{Numerical solutions of the mean-field equations of the CRP
    model with selection of specific domain classes. Left panel: cost
    function $\mathcal{F}(n)$ for different values of $\alpha$. Right
    panel: F(n) plotted in linear and logarithmic (inset) scales.}
  \label{fig:num_fitness}
\end{figure}

The above equations have the following consistency properties
\begin{itemize}
\item $\partial_n \Big(k^{+}+ k^{-}+F\Big)=0$, hence $k^{+}+ k^{-}+F =
  D\quad \forall n$. 
\item $\partial_n F \leq 1$, hence $ F(n) \leq n$. 
\item $\partial_n F \geq 0$, $\partial_n k^+ \geq 0$ and $\partial_n
  (F + k^+)  \geq 0$ so that $F$ grows faster than $k^+$ decreases.  
\end{itemize}
Choosing the initial conditions from empirical data $n_0, F(n_0)$ size
and number of domain classes of the smallest genome, we have, since
$F(n_0) < n_0$ and $\alpha \le 1$, 
\begin{displaymath}
  \frac{ \alpha F(n_0) + \theta } { n_0 + \theta } < 1 \ \ .
\end{displaymath}
It is simple to verify that under this condition the system always has
solutions that relax to a finite value $F_{\infty} < D $. Indeed, after the
time $n^*$ where $k^+(n*) = 0$, the equations reduce to $\partial_n
k^+ = 0$, $k^- = D- F$ and 
\begin{displaymath}
  \partial_n F(n)=\left(\frac{\alpha F(n)+\theta}{n+\theta}\right)^2
\end{displaymath}
immediately giving our result. 

Numerical solutions of Eq.~(\ref{eq:din}) give the same behavior for
$F(n)$ as the direct simulations (figures \ref{fig:num_fitness}A, and
figure~\ref{fig:f_n}B of the main text). In particular, while this
function grows as a power law for small genome sizes, it saturates at
the relevant scale, giving good agreement with the data.  This
behavior is connected to the finite size of the pool of universal
domain families, which we can interpret as the effect of a certain
optimality in the core functions of the different organisms.  The
internal laws of domain usage of this model were obtained from direct
simulations only, and, as discussed in the main text, give a more
quantitative agreement with the data (figure \ref{fig:hist}B of teh
main text).  Finally, one interesting point can be made about the
dynamics of the cost function. Figure \ref{fig:num_fitness}B, shows that,
for large values of $\alpha$ (above $0.7$) this function reaches a
maximum at sizes between 2000 and 4000. This is also where most of the
genomes in the data set are found, indicating that this range of
genome sizes may allow the optimal usage of universal and contextual
domain families.

\section{Other Variants of the CRP}
\label{sec:variants}

We discuss here mean-field arguments for the robustness of our results
on the asymptotics of $F(n)$ for two variants of the original model,
including a small domain loss rate and global duplications.


\paragraph{Global Duplications.}
One can consider the presence of global duplication moves. At each time
step, if duplication is chosen, a number of domains selected with
$q>1$ trials from a binomial distribution with parameter $p_O^i$ is
duplicated in the same time step. The innovation step remains the
same.
In this case, it is not possible to measure time with the size $n$ of
the genome, but this observable follows the evolution equation
\begin{equation}
  \dot{n} = q p_O + p_N \ \ ,
\label{eq:ndt}
\end{equation}
where $ \dot{\ }$ indicates the derivative with respect to time $t$.  
In terms of $t$, our mean field equations are worked out simply as
$\dot{F}(t) = p_N$ and $\dot{K_i}(t) = q p_O^i$. Using Eq.~(\ref{eq:ndt}),
  they can be simply converted in terms of $n$, yielding
  \begin{displaymath}
    \partial_n F(n)= \frac{\alpha F(n)+\theta}{q n + (q-1) \alpha F(n)
      + \theta} \ \ ,
\end{displaymath}
and
\begin{displaymath}
  \partial_n K_i(n) =   \frac{K_i-\alpha}{n+\frac{\theta}{q}} \ \ .
\end{displaymath}
The first equation gives as leading scaling $F(n) \sim n^{(\alpha /
q) } $, showing that the growth of $F$ is pushed towards effectively
lower values of $\alpha$ by global duplications, as a consequence of
the rescaling of time by the global moves. The dynamics for $K_i$,
instead, is affected only by a renormalization of the parameter
$\theta$. The qualitative results of the model are therefore stable
to  the introduction of a global duplication rate, in the hypothesis
that the extent of these duplications does not scale with $n$.

\begin{figure}[htbp]
  \centering \includegraphics[width=0.7\textwidth]{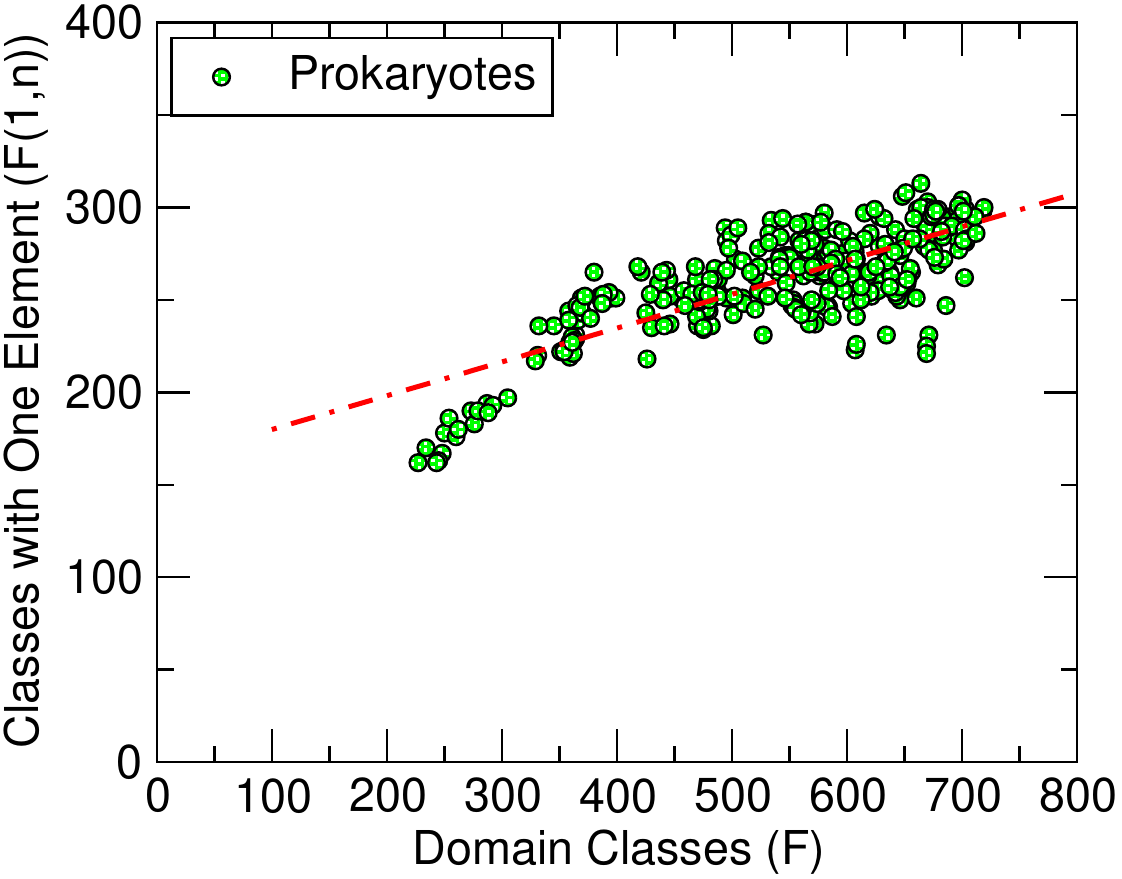} 
  \caption{Number of domain classes with one member (related to
    $F(1,n)$) from the bacteria data set for superfamilies, plotted
    as a function of the number of domain classes (realted to $F$).}
  \label{fig:F1bact}
\end{figure}

\paragraph{Domain Loss.}
A second interesting variant of the model considers the introduction
of a homogeneous domain deletion, or loss rate.  Domain loss is known
to occur in genomes. However, it is not considered in our basic model
for simplicity and economy of parameters. In order to introduce it in
the CRP, we define a loss probability $p_L = \delta$. This is equally
distributed among domains, so that the \emph{per class} loss
probability is $p_L^i = \delta \frac{K_i}{n} $. Consequently, the
duplication and innovation probability $p_O$ and $p_N$ are rescaled by
a factor $ (1-\delta)$. The mean-field evolution equation for the
number of domain classes becomes then
\begin{displaymath}
  \dot{F}(t) =  (1-\delta) \frac{\alpha F+\theta}{ n + \theta} 
               - \delta \frac{F(1,n)}{n} \ \ ,
\end{displaymath}
where the sink term for $F$ derives from domain loss in classes with a
single element, quantified by $F(1,n)$.

In order to solve this equation, one needs an expression for
$F(1,n)$. Here, we report an argument based on the fact that in the
empirical data, for large $n$, $F(1,n) = \gamma F(n) $, with
$0<\gamma<1$ (figure \ref{fig:F1bact}). This is also confirmed by
direct simulation of the model.

Using this experimentally motivated ansatz, we can
show that for small $\delta$, the scaling of $F(n)$ is subject only to
a small correction.
Again, since time does not count genome size, one has to consider the
evolution of $n$ with time $t$, given in this model simply by $
\dot{n} = 1 - 2 \delta$. Using this equation it is possible to obtain
the evolution equation for $F(n)$.  Considering an expansion in small
$\delta$ and large $n$, this reads to first order
\begin{displaymath}
   \frac{\partial_n F(n)}{F(n)}= 
   \frac{\alpha}{n} \left[ 1 + \delta \left( \frac{\alpha -
         \gamma}{\alpha} \right) \right] \ \ .
\end{displaymath}
The above equation gives the conventional scaling for $F(n)$, with the
aforementioned correction. Note that the correction could be positive
or negative, depending on the relative values of $\alpha$ and
$\gamma$.  An analogous argument holds for $\alpha = 0$.


\end{document}